\documentclass[twocolumn,amsmath,amssymb,floatfix,superscriptaddress,10pt,aps,pre]{revtex4-1}

\usepackage{graphicx}
\usepackage{amsmath,amssymb,amsfonts}
\usepackage{amsthm}
\usepackage{mathrsfs}
\usepackage{xcolor}

\newcommand{\bvec}[1]{{\mathbf{\string#1} }}

\begin{document}

\title{A Hitchhiker's Guide To Active Motion}

\author{Tobias Plasczyk}
\email{tobias.plasczyk@hhu.de}
 \affiliation{Institut f\"{u}r Theoretische Physik II: Weiche Materie, Heinrich-Heine-Universit\"{a}t D\"{u}sseldorf, D-40225 D\"{u}sseldorf, Germany}

 \author{Paul A.\ Monderkamp}
 \affiliation{Institut f\"{u}r Theoretische Physik II: Weiche Materie, Heinrich-Heine-Universit\"{a}t D\"{u}sseldorf, D-40225 D\"{u}sseldorf, Germany}

 \author{Hartmut L\"owen}
\email{hartmut.loewen@hhu.de}
 \affiliation{Institut f\"{u}r Theoretische Physik II: Weiche Materie, Heinrich-Heine-Universit\"{a}t D\"{u}sseldorf, D-40225 D\"{u}sseldorf, Germany}

\author{Ren\'e Wittmann}
\email{rene.wittmann@hhu.de}
 \affiliation{Institut f\"{u}r Theoretische Physik II: Weiche Materie, Heinrich-Heine-Universit\"{a}t D\"{u}sseldorf, D-40225 D\"{u}sseldorf, Germany}
 \affiliation{Institut für Sicherheit und Qualität bei Fleisch, Max Rubner-Institut, D-95326 Kulmbach, Germany}

 \date{\today}

\begin{abstract}
Intelligent decisions in response to external informative input can allow organisms to achieve their biological goals while spending very little of their own resources.
In this paper, we develop and study a minimal model for a navigational task, performed by an otherwise completely motorless particle that possesses the ability of \textit{hitchhiking} in a bath of active Brownian particles (ABPs).
Hitchhiking refers to identifying and attaching to suitable surrounding bath particles.
Using a reinforcement learning algorithm, such an agent, which we refer to as intelligent hitchhiking particle (IHP), is enabled to persistently navigate in the desired direction.
This relatively simple IHP can also anticipate and react to characteristic motion patterns of their hosts, which we exemplify for a bath of chiral ABPs (cABPs).
To demonstrate that the persistent motion of the IHP will outperform that of the bath particles in view of long-time ballistic motion, we calculate the mean-squared displacement and discuss its dependence on the density and persistence time of the bath ABPs by means of an analytic model.
\end{abstract}

%\keywords{Active Brownian Particles, Circle Swimmers, Hitchhiking, Persistent Motion, Machine Learning, Reinforcement Learning, Q-Learning}

\maketitle

\section{Introduction}\label{sec1}
Living organisms possess the ability to consume energy from the environment and utilise it for developing biologically advantageous strategies.
One famous example are swimming organisms which turn this energy into directed self-propelled motion \cite{marchetti2013hydrodynamics,elgeti2015physics}.
This active behaviour leads to several advantages when it comes to efficiently exploring space, finding new resources or escaping predators.
In the usual scenario, however, directed motion alone is barely a sufficient tool for most organisms when it comes to the preservation of one's livelihood.
Therefore, the archetypal survival strategy in biology incorporates at least a coupling to a rudimentary form of intelligence,
which is characterised by a sensory input and a means of interpretation \cite{jacob2004bacterial,hellingwerf2005bacterial,majumdar2017bacterial,kaspar2021rise}.
In the human world, it is utterly common that motorless individuals utilise the means of motorisation of others.
This behaviour, popularly termed hitchhiking \cite{schlebecker1958informal},  originates from the desire to optimise one's spending of energy, e.g., for economical or ecological reasons.

There are also a lot of motorless organisms in nature, such as certain species of bacteria, archaea, algae, marine plankton or atmospheric aeroplankton~\cite{smith2013aeroplankton}, which thrive despite entirely lacking the ability for self-propulsion.
For example, airborne spores and pollen  rely on wind currents \cite{markgraf1980pollen} or terrestrial dust storms~\cite{shinn2003atmospheric} as their only means of transportation.
Even the concept of human hitchhiking is found in more evolved life forms,
which developed ingenious navigational strategies that involve attaching themselves to other, usually larger, species~\cite{favet2013microbial, muok2021intermicrobial,grossart2010bacteria,siddique2021seasonal}.
Specifically, many marine organisms such as \textit{Barnacles}, \textit{Jellyella} and also microscopic bacteria can move over considerable distances, through layers of the ocean, or travelling upstream along with their hosts
\cite{grossart2010bacteria,siddique2021seasonal,howard1959predaceous,kano2009hitchhiking,vieira2015membraniporopsis}.
Other species elevate the concept of hitchhiking by forming symbiotic relationships that surpass the mere desire for transportation.
In addition to being able to swim on its own, the \textit{suckerfish} attaches to larger hosts, profiting from the protection, nutrition and fast incoming water flows, which aids its respiration, while, conversely, the host profits from the provided skincare~\cite{nicholson2021hitchhiker,norman2021three}.

 Due to the interdisciplinary importance of active motion \cite{tevrugtrevrev}, huge efforts were dedicated to theoretically characterise the self propulsion in both living matter and inanimate systems \cite{ramaswamy2010mechanics, bechinger2016active, mestre2022colloidal}.
One standard model are active Brownian particles (ABPs),
which propel themselves with a constant velocity along their instantaneous orientation that undergoes rotational diffusion.
This toolbox has stimulated theoretical predictions, ranging from analytic results on the single particle level \cite{ten2011brownian, kurzthaler2016intermediate, jeggle2020pair, caraglio2022analytic} to insight into a vast range of collective phenomena \cite{takatori2014swim, o2023introduction, caprini2020spontaneous, caprini2023entropons}.
 For a more realistic description of natural or artificial experimental systems \cite{kummel2013circular,campbell2017helical,scholz2018inertial},
several generalisations of the ABP model were devised, including
chiral ABPs (cABPs) which  tend to move in circles \cite{teeffelen2008loewen,sevilla2016diffusion,jahanshahi2017brownian,chepizhko2019ideal},
anisotropic self-propelled particles which have different angular dynamics \cite{ten2011brownian,wittkowski2012self}
and can collectively align their direction of motion \cite{bar2020self},
or active Langevin particles which incorporate the ubiquitous inertial effects from the macroscopic world \cite{lowen2020inertial,sprenger2021time}.
Most notably, more recent research avenues evoke machine learning to gain deeper physical insight into active matter systems \cite{dulaney2021machine,colen2021machine}
or explore the idea of supplementing the ABP model with machine learning tools to enhance their navigation abilities \cite{schneider2019optimal,cichos2020machine,monderkamp2022active,nasiri2023optimal,grauer2024optimizing}.
The persistent motion of intelligent motorless matter has received considerably less attention.

Here, we provide a minimal model for understanding hitchhiking in nature.
More specifically, we place a single intelligent hitchhiking particle (IHP) in a bath of ABPs. The IHP is unable to propel itself but can attach itself to the ABPs and follow their path.
To model sensory input coupled with intelligent decision-making aiming towards directed motion, this attachment is steered through reinforcement learning \cite{sutton2018reinforcement}.
The learning objective is to reach the top of the simulation box as fast as possible,
while the perceived information about the surrounding bath is limited.
Over the course of the training protocol, the IHP deduces a strategy
which allows it to  selectively attach to bath ABPs with favourable orientations.
 The final selection rule depends on the bath particles' degree of persistence,
indicating that the  IHP can balance between (i) accepting a possibly longer waiting time and only joining for highly promising travel directions (better for a highly persistent bath)
or (ii) taking the odds for a larger range of initial travel directions
(better for more erratic movement of bath particles).
The symmetry of this transport problem is broken by the possibility to let go of the travel partner
if its orientation turns out to develop unfavourably.
Moreover, the IHP learns to anticipate the circular movement of cABPs
by shifting the interval of favourable orientations according to the circular frequency of such bath particles.
Once acquired, the IHP's strategy can also be transferred to
a system of particles interacting with a soft potential.

This paper is structured as follows.
In Sec.~\ref{sec11} we describe the $Q$-learning process and our different IHP models, which we then analyse in Sec.~\ref{sec3} to discuss how the learning results and persistent motion depend on the properties of the active bath.
We conclude in Sec.~\ref{sec13}

\begin{figure}[t]
    \centering
    \includegraphics[width = 0.475\textwidth]{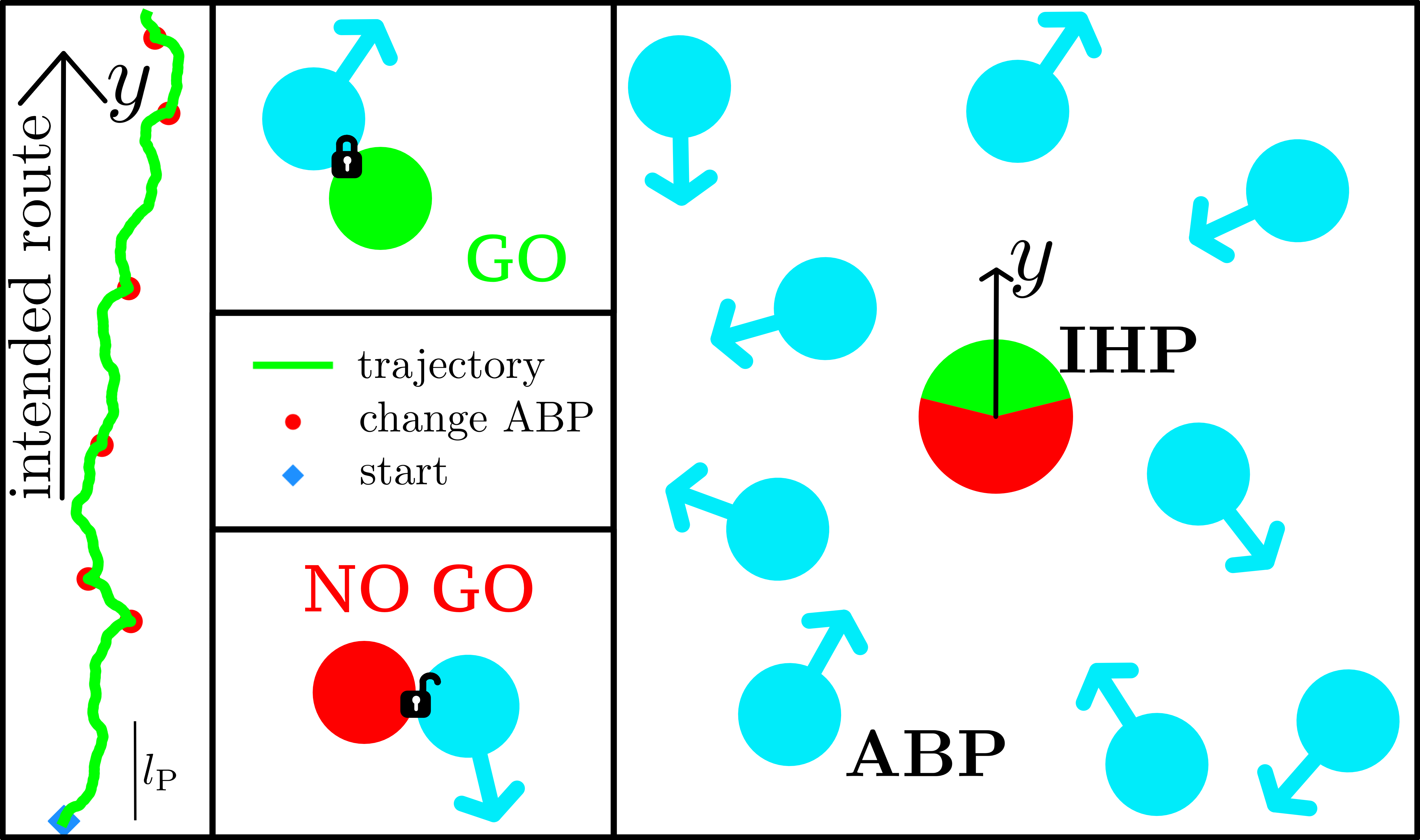}
    \caption{\textbf{Hitchhiking strategy to perform persistent Brownian motion.} An intelligent hitchhiking particle (IHP, green/red or dark/light grey) is immersed in a bath of non-interacting active Brownian particles (ABPs, cyan).
    Each ABP moves in a certain direction for their persistence length $l_\text{p}$
    (see Fig.~\ref{fig_model} for details on the model).
    The IHP learns to fulfil a navigational task by making binary decisions whether to attach and follow the closest ABP (action: GO, green or light grey) or not to attach or let go (action: NO GO, red or dark grey).
    The fully trained IHP can then persistently move to the top (green trajectory), i.e., in positive $y$-direction, on a length scale much larger than $l_\text{p}$ (see the scale bar at the bottom left), changing its travel partner if necessary (red dots).
    This decision diagram, resulting from a $Q$-learning algorithm, is displayed as a pie chart decorating the IHP, whose colour indicates the chosen action depending on the current orientation (cyan arrows) of the nearest ABP (see Fig.~\ref{fig_learning_prog_ep} for details on the learning process).}
    \label{fig_concept}
\end{figure}

\section{Models for an intelligent hitchhiking particle (IHP)}\label{sec11}

As illustrated in Fig.~\ref{fig_concept}, our model consists of a two-dimensional environment of non-interacting active Brownian particles (ABPs)
constituting potential travel partners of the intelligent hitchhiking particle (IHP).
Since it possesses no own means of propelling, its movement is governed by that of the ABP to which it temporarily attaches.
The ultimate goal is to learn a strategy, which allows the IHP to travel in the intended direction as fast as possible.
The motion pattern resulting from a successful hitchhiking strategy is exemplified in the supplementary video.

\begin{figure}[t]
    \centering
    \includegraphics[width = 0.475\textwidth]{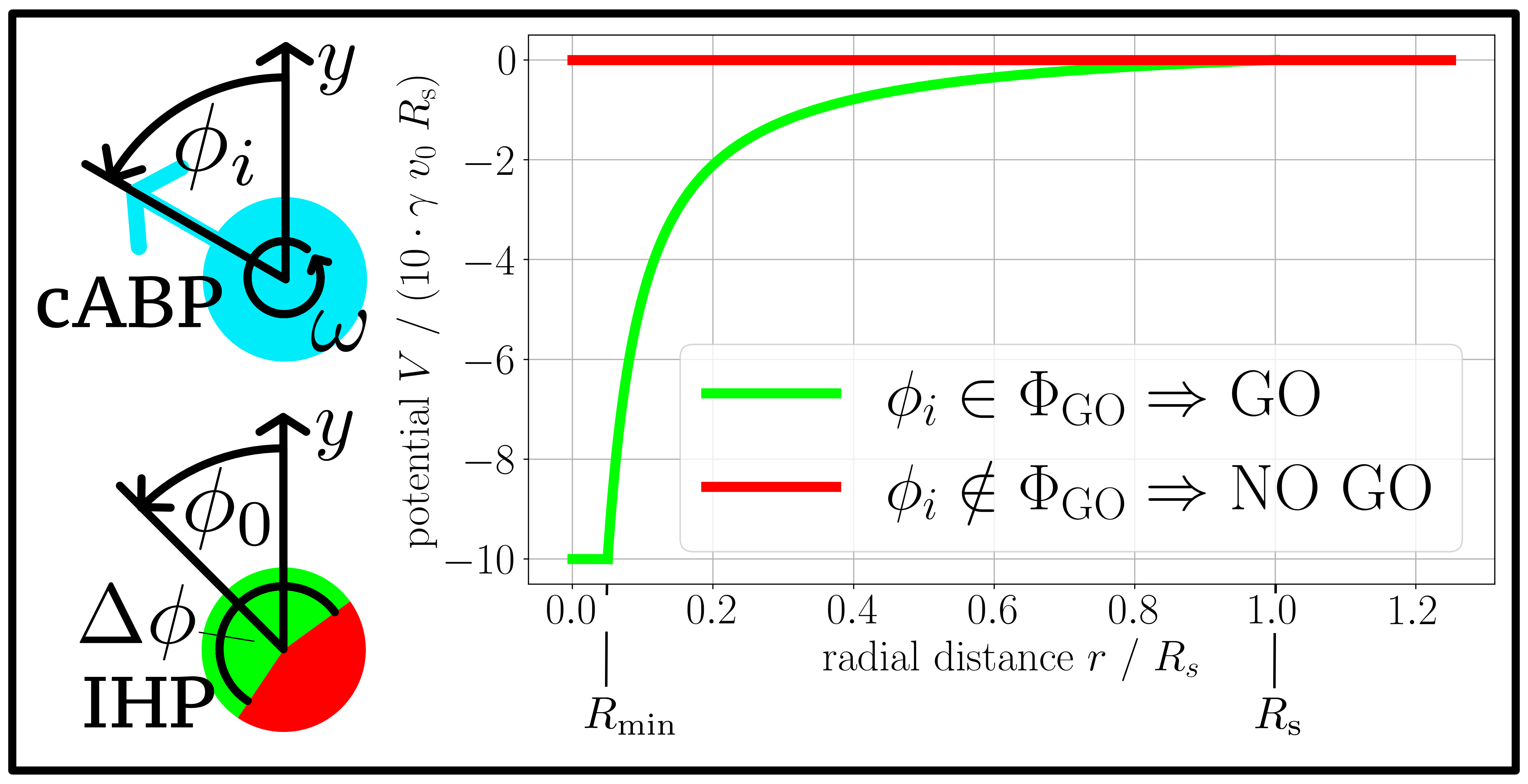}
    \caption{\textbf{Interaction of the IHP with the bath ABPs.}
    The ABPs (top left) move according to Eqs.~\eqref{eq:eq_of_motionR} and \eqref{eq:eq_of_motionPHI} with constant self-propulsion velocity $v_0$ in the direction indicated by their instantaneous orientational angle $\phi_i$, which is subject to rotational Brownian motion.
    In addition, chiral ABPs (cABPs) experience an internal torque resulting in a constant circular frequency $\omega$.
    The fully trained IHP (bottom left) is characterised by the decision diagram, indicating the values of $\phi_i$ at which it chooses the actions GO or NO GO, compare Fig.~\ref{fig_concept}.
    This internal memory is quantified by the GO-angle $\Delta\phi$, denoting the probability that a certain travel partner is chosen,
    and the anticipation angle $\phi_0$, indicating the mean orientation of the chosen travel partner.
    Hence, a GO decision is made if $\phi_i\in\Phi_\text{GO}$, as defined in Eq.~\eqref{eq_PhiGO}.
    Based on this decision diagram, we consider two models for the IHP motion.
    The first model is the $Q$-learning IHP, which shares the path of the chosen ABP according to Eq.~\eqref{eq_IHPmotion} and is used for training.
   As a second model, we transfer the acquired knowledge to the potential IHP, which has its own equation of motion~\eqref{eq_IHPmotionPOT} and is attracted to all favourable bath particles within range.
   The corresponding force is generated by the non-reciprocal interaction potential $V(r,\phi_i)$ plotted on the right.
    }
    \label{fig_model}
\end{figure}

The essence of our model is summarised in Fig.~\ref{fig_model} and details are given in the remainder of this section.
The stochastic dynamics of the bath ABPs is characterised by persistent
motion in the direction of their current orientations $\phi_i$, which we recapitulate in Sec.~\ref{sec_ABP}.
The training process, in which the IHP learns how to interact with its environment, is described in Sec.~\ref{sec_IHPqlearning}.
The central learning objective is to obtain a decision diagram, representing an interval $\Phi_\text{GO}$ of favourable angles on the unit circle,
 such that the IHP wants to attach to an ABP (denoted by GO) if $\phi_i\in\Phi_\text{GO}$ and does not want to attach or let go (denoted by NO GO) if $\phi_i\notin\Phi_\text{GO}$.
To characterise the IHP's behaviour, we thus introduce a GO-angle $\Delta\phi$, which specifies the size of this interval, and an anticipation angle $\phi_0$, indicating its location on the unit circle.
Thus, we have (always implying a $2\pi$-periodicity of the polar angle)
\begin{align}
    \Phi_\text{GO}= \left[\phi_0-\frac{\Delta\phi}{2},\phi_0+\frac{\Delta\phi}{2}\right]\,.
    \label{eq_PhiGO}
\end{align}
In Sec.~\ref{sec_IHPpotential}, we translate the decision diagram of a fully trained IHP to a (nonreciprocal) interaction potential, which only generates a force on the IHP towards an ABP if $\phi_i\in\Phi_\text{GO}$.
Both IHP models are summarised and compared in Sec.~\ref{sec_modov}.

\subsection{Active Brownian particles (ABPs) \label{sec_ABP}}

We consider a bath of $N$ non-interacting ABPs (top left drawing in Fig.~\ref{fig_model}),
labelled as $i=1,\ldots,N$, in a periodic square simulation box with side lengths $L$, where the bath density is defined as $\rho=N/L^2$.
These bath particles move at a constant self-propulsion velocity $v_\text{0}$ in the direction of their orientation vector $\bvec{u}_i(t) = (\cos(\phi_i(t)) , \sin(\phi_i(t))$ in polar coordinates.
The orientation angles $\phi_i$ are defined such that $\phi_i = 0$ corresponds to motion in the preferred travel direction of the IHP, as drawn in Fig.~\ref{fig_concept}.
They change over time due to rotational diffusion characterised by Brownian white noise $\xi_i$ with $\langle \xi_i(t) \rangle = 0 $ and $\langle \xi_i(t) \xi_j(t') \rangle = 2\delta_{ij} D_\text{r}\delta (t-t')$, where $D_\text{r}$ is the rotational diffusion constant.
Neglecting translational diffusion,
the equations of motion for the ABPs' centres-of-mass $\bvec{r}_i(t)$ read
\begin{align}
    \dot{\bvec{r}}_i(t) &= v_\text{0} \bvec{u}_i(t)
    \,\label{eq:eq_of_motionR}
\end{align}
In addition, we consider the angular dynamics
\begin{align}
    \dot{\phi}_i(t) &= \xi_i + \omega\,.
    \label{eq:eq_of_motionPHI}
\end{align}
including a constant circular frequency $\omega$, which leads to ABPs that have the tendency swim on circular trajectories.
We use these chiral active Brownian particles (cABPs)
to investigate the learning behaviour of the IHP in different environments.

The equations of motion~\eqref{eq:eq_of_motionR} and \eqref{eq:eq_of_motionPHI} are integrated, using a forward Euler-Maruyama method with a finite time step $\Delta t$.
Specifically, we discuss results upon varying the rotational diffusivity $D_\text{r}$, which sets the persistence time $\tau=D_\text{r}^{-1}$ and  length $l_\text{p} = \frac{v_\text{0}}{D_\text{r}}$ of an ABP (with $\omega=0$), and the additional circular frequency $\omega$ of a cABP, while keeping the self-propulsion velocity $v_0$ fixed.

\subsection{$Q$-learning IHP \label{sec_IHPqlearning}}

The IHP (bottom left drawing in Fig.~\ref{fig_model}) cannot move through self-propulsion.
Its only means of transportation is to attach to the nearest ABP.
To mimic realistic behaviour, only bath particles within a certain perception range,
indicated by the scan radius $R_\text{s}$, can be considered and the IHP requires a perception time $\tau_Q$ before being able to make a new decision.
In more detail, the IHP makes its $(n+1)$th decision at time $t=n\tau_Q$ with $n\geq0$.
If there are potential travel partners within the scan radius $R_\text{s}$, it picks the nearest one.
Further, if this selected ABP has a favourable orientation, the IHP will share its path for a time span of at least $\tau_Q$ (GO).
Otherwise, the IHP will rest at its current position and wait for a time span of $\tau_Q$ until the next decision  can  be made (NO GO).
Note that this also applies to a NO GO decision causing the IHP to leave its current travel partner.

We thus model the time evolution of the IHP position $\bvec{r}(t)$, which starts at $\bvec{r}(0)=\bvec{0}$, separately for each perception time span $n\tau_Q<t\leq (n+1)\tau_Q$ between two decisions according to
\begin{align}
   \bvec{r}(t) &=
   \begin{cases}
     \bvec{r}_i(t)\, & \text{if}\  \begin{cases}\phi_i(n\tau_Q) \in \Phi_\text{GO}\ \ \&\\ |\bvec{r}(n\tau_Q)-\bvec{r}_i(n\tau_Q)| \leq R_\text{s}\end{cases}\\
     \bvec{r}(n\tau_Q)\, & \text{else}
   \end{cases}\cr
   i&=\underset{j}{\arg \min} \;|\bvec{r}(n\tau_Q)-\bvec{r}_j(n\tau_Q)|\,,
    \label{eq_IHPmotion}
\end{align}
where $i$ labels the ABP closest to the IHP.
The set $\Phi_\text{GO}$ of favourable orientational angles, as given in Eq.~\eqref{eq_PhiGO}, follows from the learning process.
These equations of motion are universal to the IHP, while only the basis of decision, encoded in $\Phi_\text{GO}$, evolves while learning.

\begin{figure*}[t]
    \centering
    \includegraphics[width = 0.85 \textwidth]{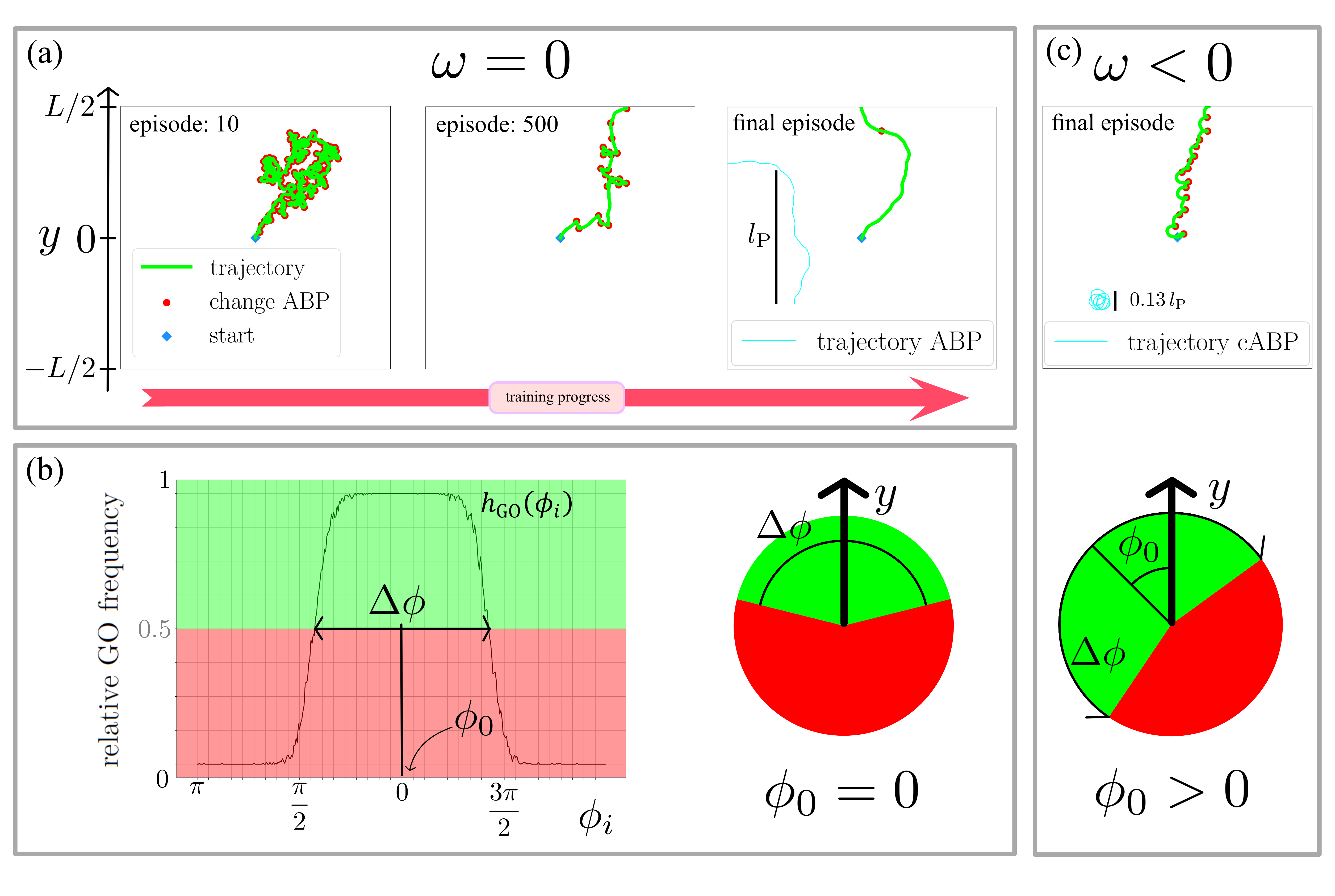}
        \caption{
        \textbf{Full learning process of the IHP.}
       \textbf{(a)}
       Trajectories (green lines)
        over the course of one training cycle, consisting of $1000$ episodes (see labels), where the IHP successively learns to navigate in a bath of ABPs with density $\rho=0.6 /R_\text{s}^2$ and rotational diffusivity $D_\text{r}=6\pi^2\cdot10^{-3}/\tau_Q$.
        The goal is to reach the top, i.e., moving half a box length $L/2=25R_\text{s}$ in positive $y$-direction.
        As in Fig.~\ref{fig_concept}, the red dots indicate the choice of action NO GO followed by a change of travel partner.
        A typical trajectory of a bath particle with persistence length $l_\text{p}=v_0/D_\text{r}$ is depicted in cyan.
        \textbf{(b)}  Obtaining the final decision rule after repeating the complete training cycle for 303 times.
        The plot shows the
        relative frequency $h_\text{GO}(\phi_i)$ of performing a GO action in a certain state, reflecting the orientation $\phi_i$ of a bath particle.
The most likely action GO (green or light grey) or NO GO (red or dark grey) for each angle is indicated in the decision diagram of the fully trained IHP (right).
        As annotated, the anticipation angle $\phi_0$ and the GO-angle $\Delta\phi$ from Eq.~\eqref{eq_PhiGO} can be directly identified in both representations.
       \textbf{(c)} According IHP and typical bath particle trajectories (top) and decision diagram of the fully trained IHP  for a bath of cABPs, which additionally have a nonzero circular frequency $\omega=-36 \pi / (100 \tau_Q)$ in their angular dynamics.
        }
        \label{fig_learning_prog_ep}
\end{figure*}

\subsubsection{Learning algorithm \label{sec_onecyclye}}

In order to enable navigation in the environment of randomly distributed ABPs, a tabular $Q$-learning algorithm is used to train
the IHP \cite{jang2019q, mahesh2020machine}.
We thus define an action space and a state space for this agent.
The action space holds all possible actions $\nu\in\{0,1
\}$ performed by the IHP,
meaning \textit{sticking to an ABP} (GO) if $\nu=1$  or \textit{not sticking to an ABP} (NO GO) if $\nu=0$.
On the other hand, the state space is given by all polar angles, i.e., the instantaneous travel direction of the selected ABP,
such that the state $\mu\in\{1,2,\ldots,360\}$ labels 360 discrete intervals.
This results in a 360$\times$2 matrix $Q_{\mu\nu}$, whose rows represent the 360 states, with each state hosting two actions, represented by the columns.
The $Q$-matrix holds the information collected during the learning process and serves as a basis of decision-making.
The resulting decision amounts to performing the action
\begin{equation}
    A_{\mu} = \underset{\nu}{\arg \max}\, Q_{\mu\nu}\in\{0,1
    \}
    \label{eq:decision}
\end{equation}
in the respective state $\mu$ that has the highest value $Q_{\mu\nu}$.
In case of equal values, $Q_{\mu0}=Q_{\mu1}$, a random action is performed.
If no ABP is within reach, no state can be defined and the agent rests until the next decision (no learning will occur in this step).

  A single training cycle consists of $1000$ episodes.
 One episode ends once the IHP reaches the top or bottom of the simulation box of fixed length $L$ or when a maximum time $T_\text{max}=6\cdot10^4\tau_Q$ is reached.
The $Q$-matrix is initialised with zeros.
As such, the agent mainly performs random actions due to their equal weight.
This reflects the lack of experience in the early stages of the learning process.
An independent probability $\epsilon$ that the IHP decides randomly decreases linearly from one to zero with increasing number of episodes.
This leaves a probability of $1-\epsilon$ to actively decide according to what the agent has learned so far, i.e., through Eq.~\eqref{eq:decision} evaluated for the current values of $Q_{\mu\nu}$.
This is to favour an exploratory approach in the beginning and an exploitative approach in the final stages of the training.

The entries $Q_{\mu\nu}$ of the $Q$-matrix are modified with  rewards or punishments.
Specifically, after the agent has performed action $\nu$ in its old state $\mu$, learning progresses according to the update formula \cite{sutton2018reinforcement}
\begin{equation}
    \label{eq:Q_learning}
    Q_{\mu\nu}^\text{new} = (1-\alpha)Q_{\mu\nu} + \alpha (R + \gamma \, \underset{{\lambda}}{\max}(Q_{\mu'\lambda}))\,
\end{equation}
which shall improve the estimate of how likely it is that this action in that state will lead to rewards in the future.
Specifically, each decision eventually leaves the agent in a new state $\mu'$ after the perception time $\tau_Q$, when a new decision has to be made.
Depending on the value $
{\max}_\lambda(Q_{\mu'\lambda})$, corresponding to the preferred action in this new state, there will be a modification on $Q_{\mu\nu}^\text{new}$,
as controlled by the \textit{discount factor} $\gamma = 0.9$.
This hyperparameter thus determines how much the values of the different states influence each other during a learning episode.
Moreover, $R$ provides a global reward (or punishment) when the agent reaches its goal at the top (or hits the bottom), such that a final update is made at the end of each episode, where we use
\begin{equation}
    R(y)=   \begin{cases}
   +100\,, & \text{if}\ y\geq L/2 \\
   -100\,, & \text{if}\ y\leq L/2 \\
   0 & \text{else}
   \end{cases} \,.
\end{equation}
Finally,  the \textit{learning rate} $\alpha = 0.01$ regulates how much of the new information is used in updating $Q_{\mu\nu}$, such that the algorithm can converge properly.

\subsubsection{
\label{sec_reptrain} Repeated training}

 To get useful results that will serve for our  later analysis of the IHP's motion,
the full learning procedure is completed only after repeating the training cycle described in Sec.~\ref{sec_onecyclye}.
We thus consider the IHP to be fully trained after having passed through $303$ independent training cycles, each consisting of $1000$ episodes.
Figure~\ref{fig_learning_prog_ep} displays this progress from the first learning steps to the final strategy, encoded in the $Q$-matrix.

In Fig.~\ref{fig_learning_prog_ep}a, we show three exemplary trajectories throughout the training  in a bath of ABPs (with $\omega=0$),
demonstrating how the strategy gradually improves.
Initially, the observed behaviour is governed by random actions, resulting in a regular change of travel partner and travel direction.
Halfway through the training,
every second decision is, on average, made on
the basis of the $Q$-matrix and the IHP increasingly benefits from the persistent ABP motion.
Eventually, it has learned to pick only upward moving travel partners and stay with them for up to about one persistence length.
The resulting trajectory displays no downward movement at all.

The main outcome of this learning algorithm is encoded in the $Q$-matrix,
which holds values for both actions in each state, where the highest value in each state defines the action, GO or NO GO, according to Eq.~\eqref{eq:decision}.
Hence, the content of the matrix elements can be condensed into a binary decision diagram only indicating the chosen action in each state.
This intrinsic memory of the IHP is represented by a pie chart, as illustrated in Fig.~\ref{fig_concept}.
However, due to our randomised setup, the final $Q$-matrix slightly differs when repeating the training of the IHP.
It is also possible to end up with a decision diagram displaying alternating actions for neighbouring states, rather than only two regions with a sharp distinction of GO and NO GO.
This is because the matrix elements for the two actions can be quite similar, in particular for travel angles corresponding to sideways ABP movement.
To obtain clean decision diagrams, we determine the relative frequency $h_\text{GO}(\phi_i)$ that the IHP makes a GO decision
after having completed $303$ independent training cycles.
As exemplified in Fig.~\ref{fig_learning_prog_ep}b,
$h_\text{GO}(\phi_i)$ is a smooth function of the ABP orientation.
We then determine the decision diagram of the fully trained IHP by associating all angles with $h_\text{GO}(\phi_i)>0.5$ and $h_\text{GO}(\phi_i)<0.5$ to GO and NO GO, respectively.
This final result can then directly be translated into the set $\Phi_\text{GO}$ of GO-angles from Eq.~\eqref{eq_PhiGO}.
As the average decision value for the APB bath is typically symmetric around $\phi_i=0$, there is no anticipation angle,
$\phi_0=0$, while the GO-angle $\Delta\Phi$ spans a range of favourable travel directions typically pointing upwards.

The flexibility of the learning procedure is verified by considering $\omega\neq0$, i.e., a bath of cABPs.
Also in this case, the trajectory of the IHP at the end of its training is directed upwards, as shown in the top of Fig.~\ref{fig_learning_prog_ep}c.
Since the characteristic radius of the circular trajectories is much smaller than the persistence length of the straight ABPs, this directed hitchhiking requires much more proactive intelligent decision making and an appropriate timing to change the travel partner.
This results in a strategy adjusted to the dynamics of the travel partner,
as reflected by the non-zero anticipation angle $\phi_0$ in the decision diagram at the bottom of Fig.~\ref{fig_learning_prog_ep}c.
In particular, $\phi_0$ typically has the opposite sign as the circular frequency $\omega$ of the bath particles
and thus conveniently quantifies how much the IHP needs to anticipate future changes in travel direction.

\subsection{Potential IHP}\label{sec_IHPpotential}

The motion~\eqref{eq_IHPmotion} of the $Q$-learning IHP, as introduced in Sec.~\ref{sec_IHPqlearning}, follows a simple identification rule,
namely travelling along with a bath ABP or not.
Its decision strategy derived from the training scheme in Fig.~\ref{fig_learning_prog_ep} is purely characterised by whether the orientation of the closest bath particle lies within a certain angular interval.
This behaviour can be imitated by purely physical interaction via a relatively simple attractive potential.
To this end, we introduce the potential IHP as an alternative model
by translating the GO action to an attractive force which pushes it towards all sufficiently close ABPs with a promising travel direction.
These directions are inferred from the decision diagram obtained in the previously completed learning processes of the $Q$-learning IHP.
Again, the motion of the bath ABPs is not affected by the presence of the potential IHP,
which means that we introduce a non-reciprocal hitchhiking interaction.

Specifically, the potential IHP  moves according to the overdamped equation of motion
\begin{align}
    \dot{\bvec{r}}(t) &= -\gamma^{-1} \sum_{i=1}^N\nabla V(|\bvec{r}-\bvec{r}_i|,\phi_i)
    \label{eq_IHPmotionPOT}
\end{align}
induced by each nearby bath particle, where $\gamma$ is the friction coefficient.
The orientation dependence of the non-reciprocal hitchhiking potential $V(r,\phi)$ follows from the binary distinction
\begin{equation}
    V(r,\phi) = \begin{cases}
     V_\text{0}(r)\,, &  \phi \in \Phi_\text{GO} \\
     0\,, &  \phi \notin \Phi_\text{GO} \\
   \end{cases}\,,
    \label{eq:potential}
\end{equation}
selecting all favourable travel directions $\Phi_\text{GO}$ of a bath particle, given by Eq.~\eqref{eq_PhiGO}.
Moreover, the radial part within $V(r,\phi)$ is described by
\begin{equation}
    V_\text{0}(r) = \begin{cases}
     5 \gamma v_0 R_\text{s}\left(1-\frac{R_\text{s}}{r}\right)\,, &  R_\text{min} \leq r \leq R_\text{s} \\
     0\,, & \text{else} \\
   \end{cases}\,.
    \label{eq:v_0}
\end{equation}
To prevent a divergence of the force, we introduce a lower cut-off to the potential, given by the minimal hitchhiking distance $R_\text{min}=R_\text{s}/20$,
while the upper perception limit is given again by the scan radius $R_\text{s}$.
This hitchhiking potential is shown in Fig.~\ref{fig_model} together with an illustration of the IHP's decision diagram.
In practice, it can happen that the potential IHP  overtakes the ABP when updating the position according to Eq.~\eqref{eq_IHPmotionPOT}
because of the finite time step $\Delta t$ in our simulations.
To prevent this discretisation artefact, the resulting displacement vector is renormalised to the distance between potential IHP  and ABP whenever it is larger (while the direction of the displacement remains the same).

\subsection{Model overview \label{sec_modov}}

Overall, we devise two different models for an intelligent hitchhiking particle (IHP), each describing individual hitchhiking capabilities.

The $Q$-learning IHP, on the one hand, is limited by its perception, as it can only make one cognitive operation of a certain complexity at discrete time steps.
It also only focuses on a single bath particle, but, if this one turns out to be a suitable hitchhiking partner, will always rigorously follow its path, irrespective of the initial distance (within the scan radius $R_\text{s}$).
This motion will be maintained until the next decision can be made after the perception time $\tau_Q$, even if the trajectory takes an unfavourable direction before.
This possibility must be appropriately anticipated in the learning process.

The potential IHP, on the other hand, is continuously pulled by all suitable travel partners (within range) at the same time instead of reacting by instantaneously attaching to a certain bath particle.
Moreover, it does not feature a discrete perception time, which brings about a strength and a weakness at the same time.
While it is less likely to be manoeuvred into an unfavourable travel direction during hitchhiking, detaching too early may increase the time it has to wait at rest.
 Keeping in mind that the decision diagram of the potential IHP is obtained from training the $Q$-learning IHP, it is not obvious which model shows the better performance in how far these features are advantageous or disadvantageous.

In the following sections,
we use $\tau_Q$ and $R_\text{s}$ as time and length scales, respectively and consider a fixed self-propulsion velocity $v_0=3R_\text{s}/(2\tau_Q)$ of the bath particles. The numerical time step is given as $\Delta t=\tau_Q/6$ and the length of the simulation box is $L=25R_\text{s}$.
The rotational diffusivity $D_\text{r}$, circular frequency $\omega$ and density $\rho$ of the bath are variable,
while learning takes place at a fixed density $\rho=0.6 / R_\text{s}^2$.

\begin{figure}[t]
    \centering
    \includegraphics[width = 0.475\textwidth]{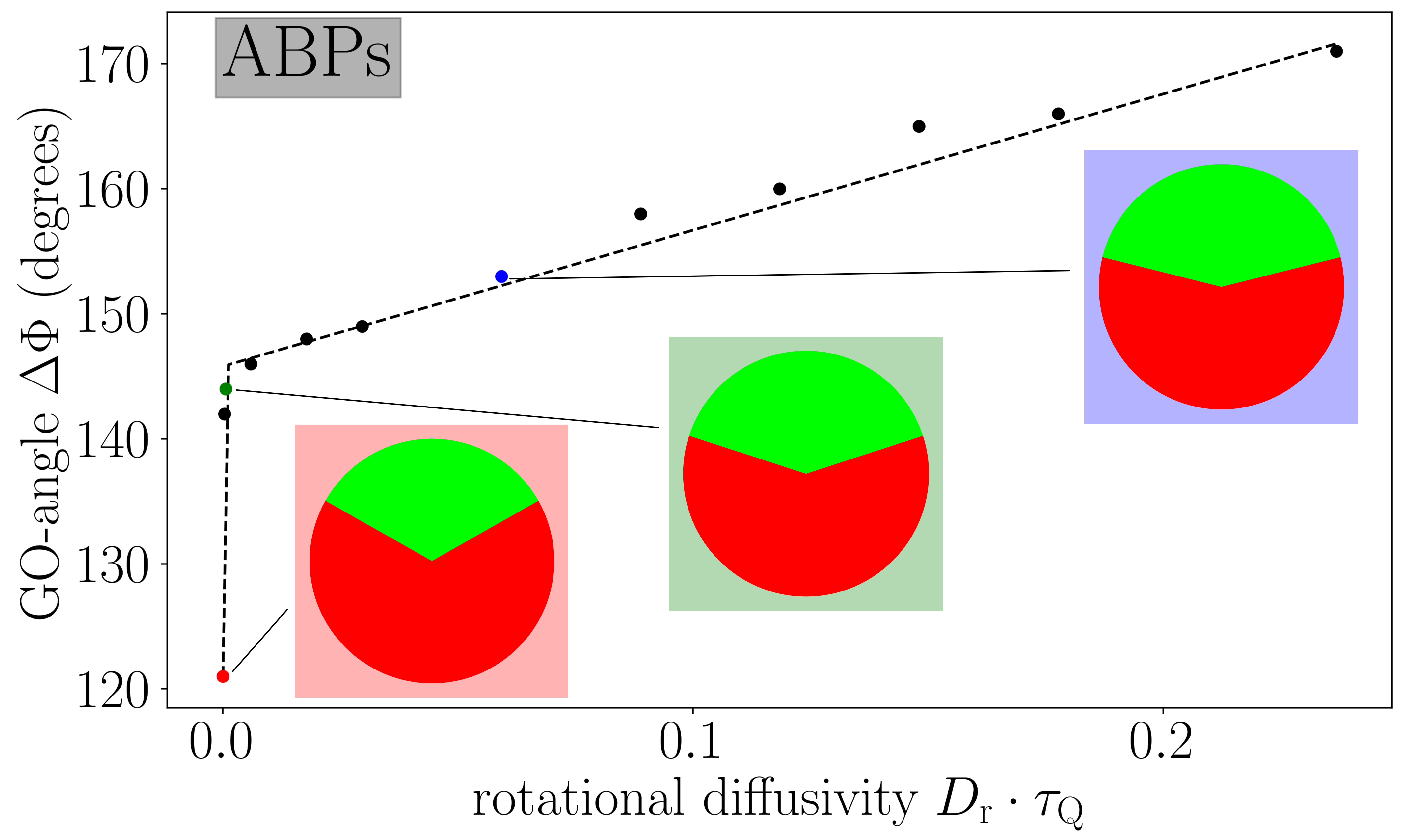}
    \caption{\textbf{Learning results
    in a bath of ABPs.}
    The GO-angle $\Delta\Phi$ of a fully trained IHP (compare Fig.~\ref{fig_learning_prog_ep})
   is shown as a function of the rotational diffusivity $D_\text{r}$ of the bath particles.
   We roughly observe a linear trend except for very small $D_\text{r}$ (highly persistent ABPs),
   where the learning environment is too small compared to the persistence length.
   The dashed line thus indicates a fit to $a+bD_\text{r}+c/D_\text{r}$ as a guide to the eye.
   The anticipation angle $\phi_0=0$ is zero in all cases, as the circular frequency $\omega=0$ vanishes for this bath.
   We also show the three decision diagrams corresponding to the values of $D_\text{r}$ (coloured points) used in our later plots.
   }
    \label{fig_delta_phi_vs_D_r}
\end{figure}

\section{Hitchhiking Behaviour}\label{sec3}

Equipped with the decision diagram obtained via $Q$-learning,
the IHP is capable of navigating in different types of baths to persistently move towards its goal (in our case in positive y-direction).
Below, we discuss its physical properties, depending on different models and bath parameters.
Specifically, we discuss in Sec.~\ref{sec2} how the decision diagrams depend on the persistence time and circular frequency of the bath particles,
before analysing the $Q$-learning IHP's mean-squared displacement in Sec.~\ref{subsec3_2}.
Finally, we compare the motion to the potential IHP in Sec.~\ref{sec_potential}.

\begin{figure*}[t]
    \centering
    \includegraphics[width = 0.95\textwidth]{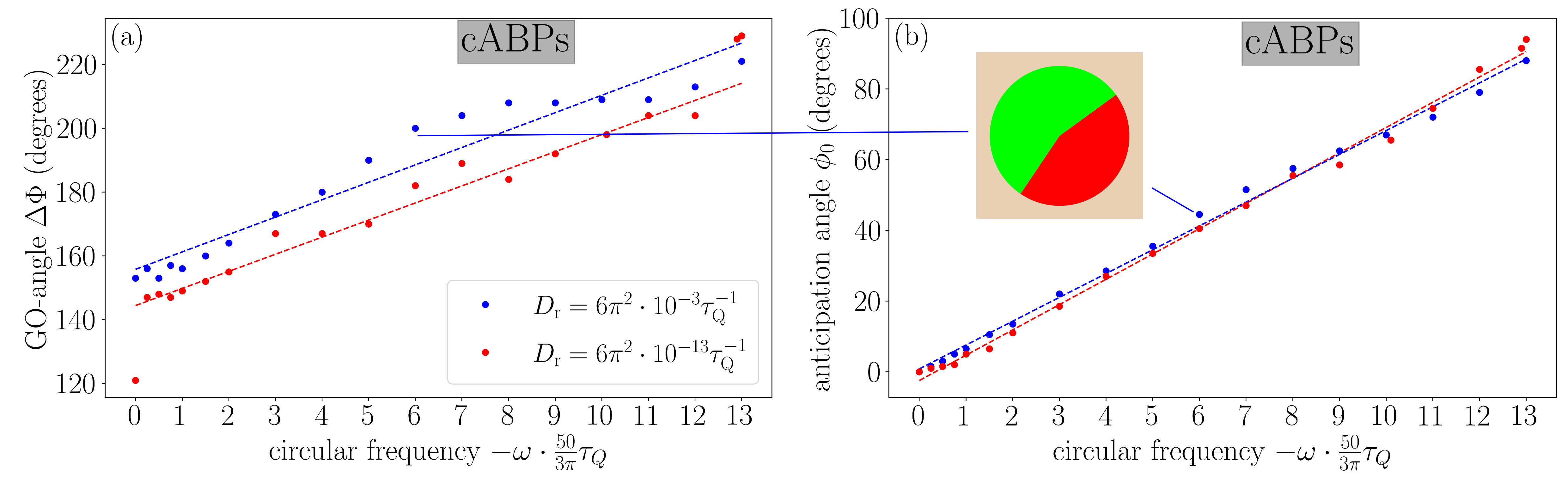}\hfill
    \caption{\textbf{Learning results
    in a bath of cABPs.}  \textbf{(a)}
The GO-angle $\Delta\Phi$ and \textbf{(b)} the anticipation angle $\phi_0$ of a fully trained IHP (compare Fig.~\ref{fig_learning_prog_ep})
   are shown as functions of the circular frequency $\omega$ of the bath particles.
We consider two rotational diffusivities $D_\text{r}$ following the colour code used in Fig.~\ref{fig_delta_phi_vs_D_r} for achiral ABPs (corresponding to the data for $\omega=0$).
   The dashed lines indicate linear fits to guide the eye (the red point for $\omega=0$ was excluded for this purpose).
   We also show the decision diagram corresponding to the parameters $D_\text{r}=6 \pi^2\cdot 10^{-3} \tau_\text{Q}^{-1}$ and $\omega=36 \pi \cdot 10^{-2} \tau_\text{Q}^{-1}$ used in Fig.~\ref{fig:IHP_vs_Pot}.}
    \label{fig_delta_phi_vs_omega}
\end{figure*}

\subsection{Learning Results}\label{sec2}

At the end of its training, the IHP has learned to reach its goal by intelligently switching travel partners.
As summarised in Fig.~\ref{fig_learning_prog_ep}, the fully trained IHP attaches to an ABP with a promising
orientation and holds on until the polar travel angle falls outside a certain range characterised by the anticipation angle $\phi_0$ and the GO-angle $\Delta\phi$.
We thus observe persistent trajectories closely resembling that of the bath particles, while a change of travel partner is observed as soon as the trajectory tends to deviate too much from the intended travel route.
In a bath of cABPs, such a change occurs more frequently at the same $D_\text{r}$, as the circular motion reduces the persistence and needs to be anticipated by the IHP.
Hence, the IHP's trajectory is characterised by sequential semicircles
and its decision diagram is asymmetric due to a nonzero anticipation angle.
Below, we quantify these learning results in more detail.

Let us first focus on a bath of ABPs with vanishing circular frequency $\omega=0$ and
investigate the dependence of the GO-angle $\Delta\phi$ on the rotational diffusivity $D_\text{r}$ or, equivalently, on the persistence time $\tau=1/D_\text{r}$.
As shown in Fig.~\ref{fig_delta_phi_vs_D_r}, $\Delta\phi$ increases  with increasing $D_\text{r}$,
which indicates that the training of the IHP has increased its awareness towards the choice of travel partner by considering a smaller range of favourable angles when the bath particles become more persistent.
In this case, the learned strategy to accept a longer waiting time for a bath particle with an optimal travel direction will naturally pay off in the long run.
 This trend can be intuitively explained by considering the consequences of choosing a travel partner whose orientation initially points sidewards.
For a highly persistent bath, the travel direction will remain the same for an extended amount of time, such that the IHP has to stick to its decision without getting an actual benefit, while better alternatives could have been available in the mean time.
For lower bath persistence, the travel direction randomises earlier, becoming either more favourable or triggering a NO GO decision allowing to  search for a better partner.
Over a large range of $D_\text{r}$, the change of GO-angle is well described by a linear relation.
A significant deviation from this trend is only observed in the extremely persistent case (small $D_\text{r}$), such that $l_\text{p}\gg L$, where
a change of a travel partner is almost never required during training.

To underline the flexibility of our $Q$-learning procedure, we now consider a bath of cABPs, for which we investigate the influence of the circular frequency $\omega$ on the decision diagram in Fig.~\ref{fig_delta_phi_vs_omega}.
As shown in Fig.~\ref{fig_delta_phi_vs_omega}a, the GO-angle $\Delta\phi$ increases approximately linearly for increasing $\omega$, since the
circular trajectories of the bath particles lead to less persistent motion (compare Fig.~\ref{fig_learning_prog_ep}c).
This observation is thus analogous to that in Fig.~\ref{fig_delta_phi_vs_D_r} for increasing the rotational diffusivity $D_\text{r}$, as discussed in the previous paragraph.
Moreover, the slope of  $\Delta\phi$ as a function of $\omega$ remains nearly independent of $D_\text{r}$, such that we expect the effect of varying $D_\text{r}$ for cABPs to be the same as for ABPs.
As in Fig.~\ref{fig_delta_phi_vs_D_r}, the only exception is for an extremely persistent bath ($D_\text{r}=6 \pi^2\cdot 10^{-13} \tau_\text{Q}^{-1}$ and $\omega=0$).

Crucially, the circular frequency of the bath cABPs also challenges the IHP to anticipate the circular movement.
This is reflected in our learning results, shown in Fig.~\ref{fig_delta_phi_vs_omega}b, by the increase of the absolute anticipation angle $|\phi_0|$ with increasing absolute circular frequency $|\omega|$, while the sign is the opposite.
 In other words, the IHP preferably selects travel partners whose orientations are not instantaneously pointing towards the goal, but will turn accordingly during the hike.
Quantitatively, the linear fit $\phi_0\approx-0.64\omega\tau_Q$ (measured in radians) tells us that the magnitude of the anticipation angle corresponds to about half the angle a bath particle turns during one perception time $\tau_Q$.
This farsighted anticipation required for high circular frequencies is thus learned successfully.
We also find that the rotational diffusion barely affects the anticipation angle, as these orientational fluctuations average out while learning to cope with this deterministic effect.

As learning only takes place if a bath particle is within range, such that a certain action can be consciously chosen, the final decision diagrams are independent of the bath density $\rho$.
Thus, the only effect of changing $\rho$, so far, is on the slope of the learning curve.
Our chosen density for the learning process yields an optimal balance between the expected number of time steps required to complete the training (which decreases for larger $\rho$) and the computational cost to evaluate each time step (which increases for larger $\rho$).
The performance of a fully trained IHP, however, very well depends on the bath density, as we examine below.

\begin{figure*}[t]
    \centering
    \includegraphics[width = 0.475\textwidth]{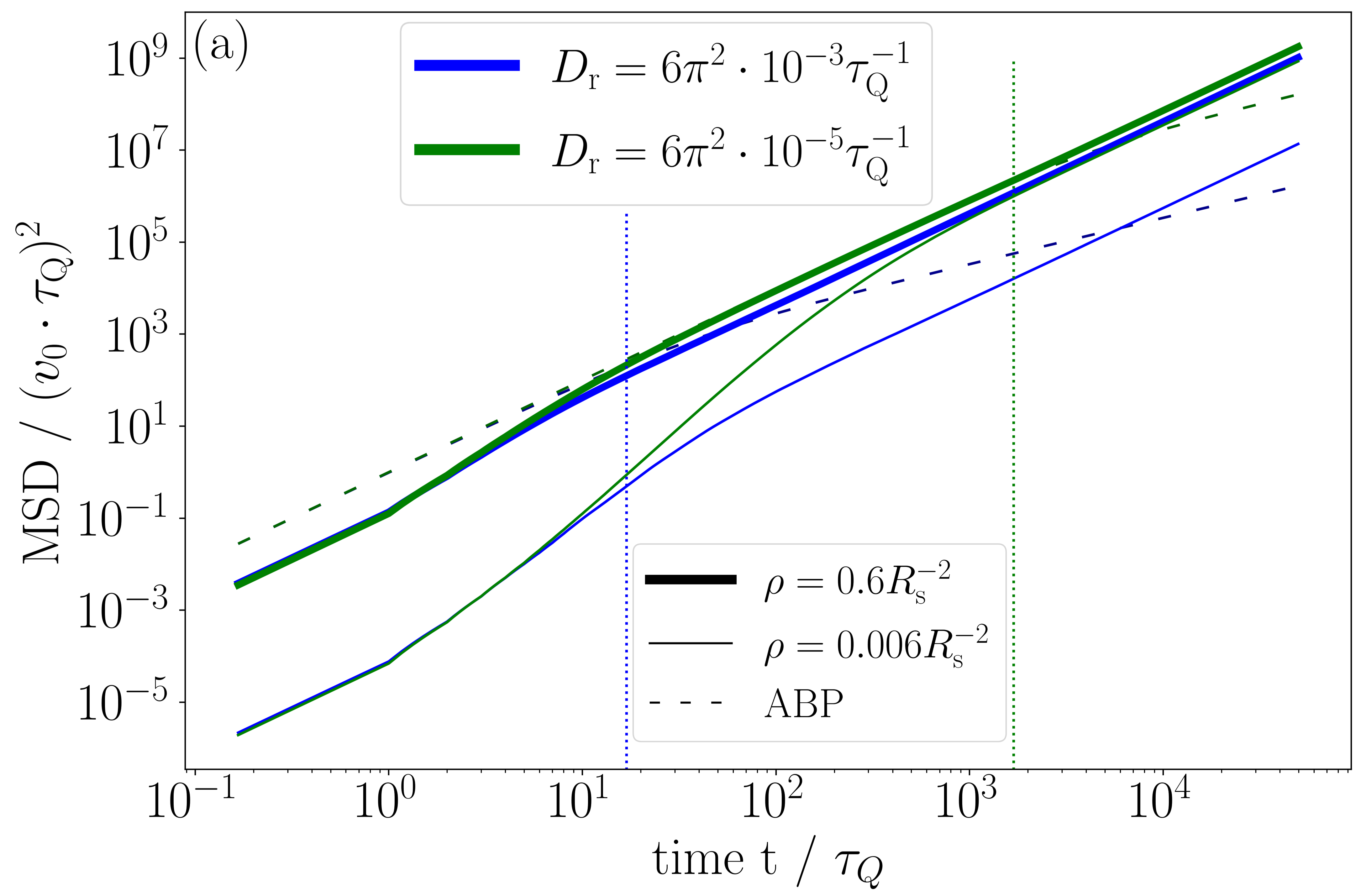}\hfill
    \includegraphics[width = 0.475\textwidth]{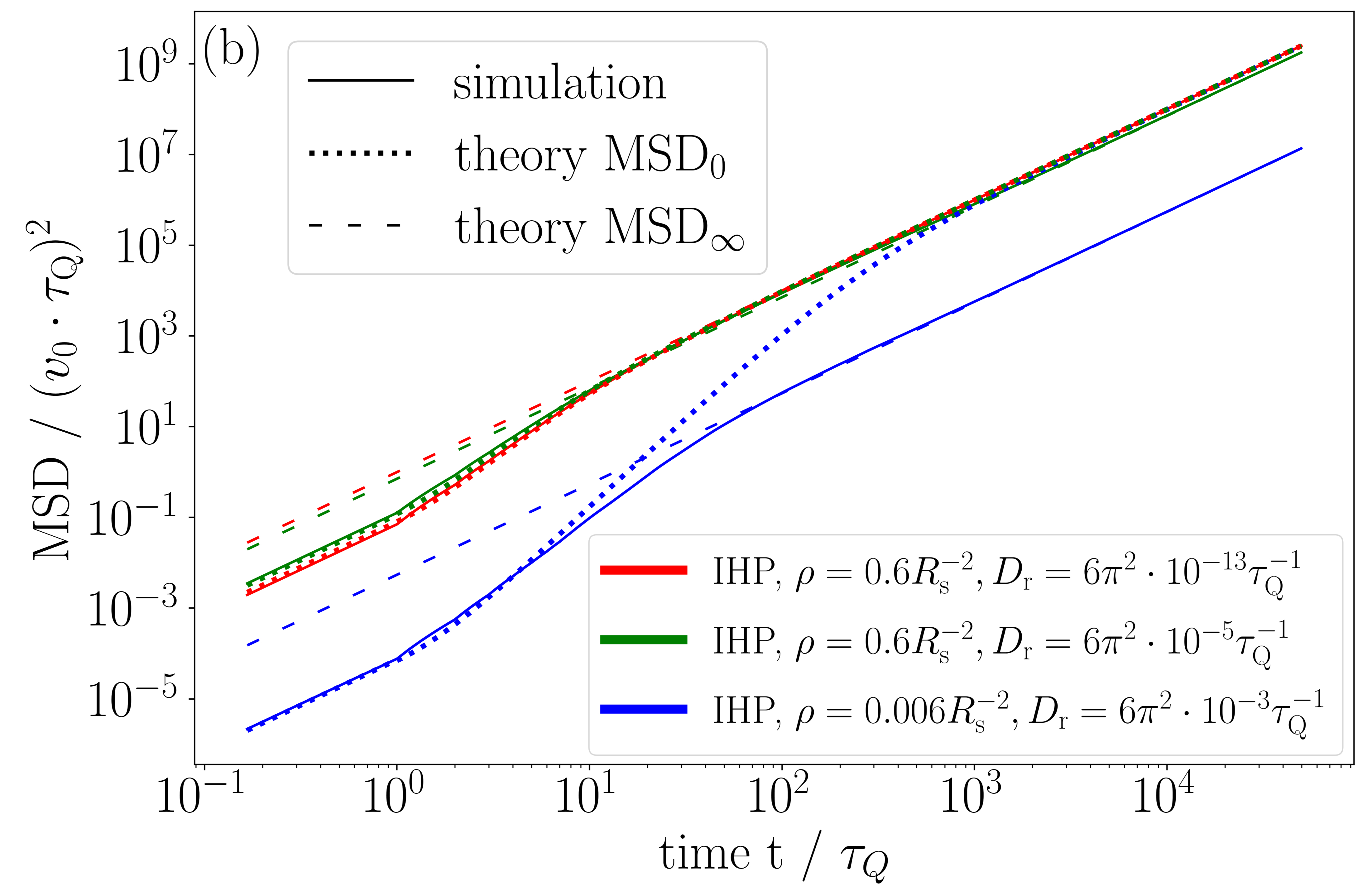}
    \caption{\textbf{Persistent motion of the $Q$-learning IHP.}
    We show the MSD for different bath densities $\rho$ and rotational diffusivities $D_\text{r}$ (solid lines with colours given in the respective legends),  where each curve is averaged over 1000 simulation runs.
    The numerical results are compared to
    \textbf{(a)} the MSD of a singlke bath ABP (dashed lines), where the vertical dotted lines indicate the persistence times $\tau=1/D_\text{r}$, and
    \textbf{(b)} the theoretical predictions derived in appendix~\ref{app_MSDTheory}, i.e.,
    the short-time approximation MSD$_0$ (dotted lines) and the long-time asymptotic behaviour MSD$_\infty$ (dashed lines).
    }
    \label{fig_MSD}
\end{figure*}

\subsection{Ballistic Motion of the IHP}\label{subsec3_2}

To illustrate the IHP's
travel characteristics, we examine its mean-squared displacement (MSD)
\begin{equation}
\text{MSD} := \left\langle  |\mathbf{r}(t)-\mathbf{r}(0)|^2  \right\rangle
\end{equation}
over the course of time.
As discussed in appendix~\ref{app_MSDTheory} (and described in the theory outlined below), the IHP can actually be well described as performing a drift motion with a time-dependent hitchhiking velocity $v_\text{H}(t)$.
Nevertheless, we consider the MSD instead of the simpler mean displacement to better assess the IHP's performance in comparison to the bath ABPs as a reference.

As shown in Fig.~\ref{fig_MSD}, the IHP has successfully learned to enable directed motion through hitchhiking, which can be seen from the ballistic long-time behaviour.
Specifically, for all parameters (bath density and rotational diffusivity), the IHP will eventually outperform a single bath ABP, whose motion becomes diffusive at times $t>\tau$ due to its finite persistence time $\tau$, compare Fig.~\ref{fig_MSD}a.
Upon closer inspection, we observe up to four different dynamical regimes, starting ballistically at very short times, followed by intermediate super-ballistic and sub-ballistic behaviour, until the IHP eventually enters its final ballistic state with a larger average velocity than in the first ballistic regime.
The most important control parameter for this dynamical behaviour and also the overall hitchhiking efficiency is the bath density.
In general, the MSD is always larger when the density is higher, as the average waiting time for a new travel partner decreases.
This effect is nearly independent of the bath particles' rotational diffusivity at short times.
The long-time performance, however, is crucially reduced if the bath is only weakly persistent and its density is low,
because of the combined effect of the IHP leaving its travel partner more often and the longer waiting time for a new one.
In contrast, hitchhiking in a dilute but persistent bath is only slightly less efficient in the long run,
while bath persistence is generally not a relevant factor in a dense bath, as there is practically always an ABP within reach.

\begin{figure*}[t]
    \centering
    \includegraphics[width = 0.975\textwidth]{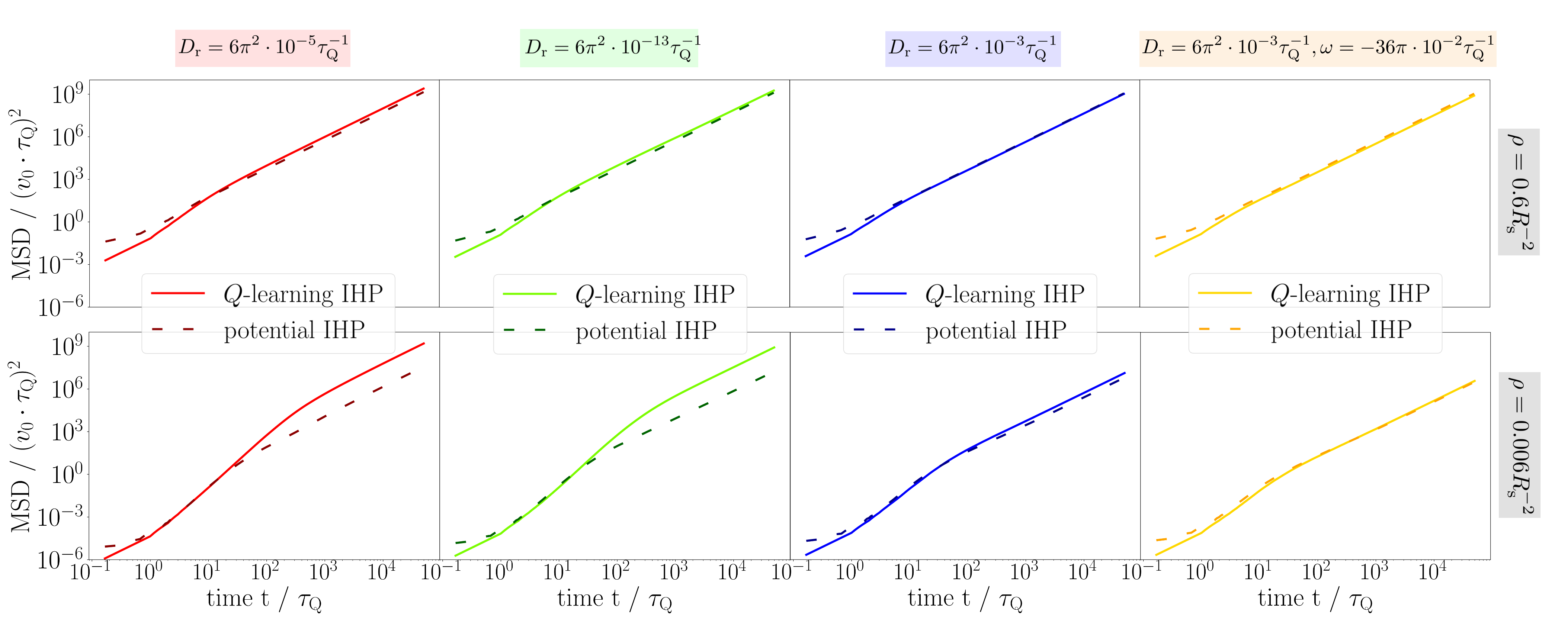}
    \caption{\textbf{$Q$-learning IHP vs.\ potential IHP.}
    We show the MSD for different bath densities $\rho$, rotational diffusivities $D_\text{r}$ and circular frequencies (as labelled) for the $Q$-learning IHP (solid lines) and the potential IHP (dashed lines).
    Each curve is averaged over 1000 simulation runs.
    }
    \label{fig:IHP_vs_Pot}
\end{figure*}

Both the qualitative behaviour and the magnitude of the MSD can be quantitatively explained via explicit considerations of the probability that a favourable travel partner is available to the IHP  at time $t$.
Our theoretical treatment is detailed in appendix~\ref{app_MSDTheory} and outlined below.
Let us first recall from Eq.~\eqref{eq_IHPmotion} that the IHP will generally either remain at rest or travel with velocity $v_0$.
Therefore, in the early stages, the IHP needs to wait to find its initial travel partner before being able to move at all.
The first regime is thus at $t\leq\tau_Q$, for which we observe ballistic motion with the average velocity $p(0)v_0$.
Given $N=\rho L^2$ bath particles in a periodic box of side length $L$,
the required probability
\begin{equation}
    p(0) = \left( 1-\left( 1-\frac{R_\text{s}^2 \pi}{L^2} \right)^{\rho L^2}\right)\frac{\Delta\phi}{2\pi}
    \label{eq:p0}
\end{equation}
to find a suitable travel partner in the IHP's observation window $R_\text{s}^2\pi$
instantaneously at $t=0$ can be exactly
derived from the binomial distribution and the fraction $\Delta\phi/(2\pi)$ of GO-angles.
If none is found, the IHP has a new chance after
every passing of its perception time $\tau_Q$,
which allows us to construct the cumulative probability $p(t)$
to having found the initial travel partner before time $t$.
As, by its nature, $p(t)$ increases as a function of time, the average hitchhiking velocity $v_\text{H}(t)=p(t)v_0$ is subject to an effective acceleration, such that the MSD becomes super-ballistic in the second regime $\tau_Q<t\ll\tau$.
As the orientation of an ABP randomises after their characteristic persistence time $\tau$, the IHP eventually needs to let go its first companion.
This may result in an intermediate deceleration on average, i.e., an effective sub-ballistic MSD regime around $t\approx\tau$ (whether this regime is visible will depend on the form of $p(t)$, which is mainly determined by $\rho$).
Hence, our prediction $p(t)$ provides an upper bound for the probability that the IHP currently has a suitable travel partner at time $t$, which is the focal point of our theoretical description.
To account for the letting go and the subsequent quest for a new travel partner,
we notice that, after this procedure has been repeated several times, a good estimate for the long-time hitchhiking velocity in the final ballistic regime for $t\gg\tau$  can be obtained by averaging $p(t)v_0$ over one persistence time of the bath particles.
Clearly, this predicted average is larger than $p(0)v_0$ in the first ballistic regime.
As indicated in Fig.~\ref{fig_MSD}b, the theoretical predictions based on the upper bound and the long-time behaviour provide an excellent representation of the simulated MSD in the corresponding time regimes.

\subsection{Comparison to the Potential Model}\label{sec_potential}

Having understood the MSD of the $Q$-learning IHP, we now turn to the potential IHP and compare the MSD from both models in Fig.~\ref{fig:IHP_vs_Pot}.
In general, we find that the potential IHP has a higher mobility at short times, while the displacement in the initial regime is rather diffusive than ballistic.
In later stages, the qualitative behaviour of both models is similar, i.e., the motion is effectively accelerated due to the increasing hitchhiking probability and eventually becomes ballistic, reflecting a successful hitchhiking strategy with persistent trajectories.
As the journey proceeds, the $Q$-learning IHP catches up and prevails over the potential IHP in the long run for most of the parameters.

Recalling the comparison in Sec.~\ref{sec_modov},
we identify two main reasons for the differences between the two models observed in the MSD.
First, as detailed in appendix~\ref{app_MSDTheoryPOT}, the probability $p(0)$ to find a suitable (initial or new) travel partner is always larger in the potential model, as the $Q$-learning  IHP can only focus on a single bath particle.
This explains the initial advantage of the potential IHP.
Second, while the $Q$-learning IHP can only make discrete decisions separated by the perception time $\tau_\text{Q}$,
a suitable bath particle always interacts with the potential IHP, which leads to  a continuous update of the instantaneous hitchhiking velocity $-\gamma^{-1}\nabla V$.
More specifically, the potential IHP
gets pulled towards the current position of the ABP, which itself will move with $v_0$ in a possibly different direction.
Assuming that this ABP will still have a favourable orientation, there are three possible scenarios for the IHP motion in the next time step:
(i) the pulling force was not directed properly, such that the ABP is out of range and the IHP
needs to wait in rest for another travel partner;
(ii) its distance to the ABP has changed and the next pull results in a different hitchhiking velocity or
(iii) the IHP ends up within a distance smaller than $R_\text{min}$ and thus does not need a kick to remain attached to the ABP, such that it remains at rest to be kicked again in the next time step.
The more random nature of the individual hitchhiking velocity (ii,iii) could be responsible for the rather diffusive motion in the initial hitchhiking phase.
Moreover, the higher susceptibility of the potential IHP to let go of its travel partner prematurely explains the catching up of the $Q$-learning IHP.
A specific scenario which could trigger this situation (i) is due to the joint attraction of several suitable travel partners  pushing the   IHP  in an average direction in between such that it will loose connection to all of them.
Such a premature detachment (compared to the $Q$-learning IHP) from a single travel partner can also be caused by the condition $\phi_i\in\Phi_\text{GO}$ being evaluated at each time step (instead of each $\tau_Q$).

In how far the initial advantages of the potential model will remain in the long run depends on different parameters, as compared in Fig.~\ref{fig:IHP_vs_Pot}.
We see that the relative long-time performance, i.e., the final average hitchhiking velocity, of the $Q$-learning IHP surpasses that of the potential IHP for a more persistent and dilute bath.
In contrast, the premature detachment of the potential IHP is less
disadvantageous both in a bath with low persistence, where the travel partner needs to be changed more frequently anyway (compare the two trajectories in Figs.~\ref{fig_learning_prog_ep}a and c),
and in a dense bath, where many suitable new travel partners are present.
To better understand the crucial role of density, let us mimic the effects described above by assuming a smaller effective scan radius of the potential IHP (mimicking the effectively looser connection).
Following appendix~\ref{app_MSDTheoryPOT}, the probability $p(0)$ to find a suitable travel partner
is then, indeed, smaller for the potential IHP in the limit of a dilute bath, while we find $p(0)\rightarrow1$ for very high densities, irrespective of the (effective) scan radius.
In contrast, the limit $p(0)\rightarrow \Delta\phi/2\pi$ of Eq.~\eqref{eq:p0} is set solely by the GO-angle of the $Q$-learning IHP, as its focus is limited to a single bath particle.

Finally, let us recall that the $Q$-learning IHP has been explicitly trained  performing its characteristic dynamics, while the  potential IHP  model requires a transfer of this knowledge into a more realistic physical model.
In this light, the performance of the latter is quite remarkable, in particular in a bath of cABPs, where its MSD is larger for all times.

\section{Conclusions}\label{sec13}

In conclusion, we have modelled  an intelligent hitchhiking particle (IHP) that gets transported by attaching to and detaching from neighbouring active particles.
It learns the rules for optimal transport by a reinforcement learning algorithm,
 demonstrating that  artificial intelligence allows to efficiently generate  directed motion from an undirected bath.
We have developed a compact analytical framework for describing the drift velocity of the IHP
and verified by numerical exploration of its
mean-squared displacement (MSD) that the persistence of this motion exceeds that of the bath particles for long times.
This ratchet-like behaviour \cite{hanggi2005brownian,di2010bacterial,reichhardt2017ratchet} of the IHP is reminiscent of the motility-induced drift motion of anisotropic tracers surrounded by ABPs \cite{kaiser2014transport,mallory2014curvature,knevzevic2020effective,granek2022anomalous}.

The learned strategy of the IHP depends on two factors.
First, the overall bath persistence is reflected by the GO angle.
 If the bath is highly persistent (small rotational diffusivity and small circular frequency), then it is advantageous to be more selective (small GO angle), as the presumed long travel time with a chosen partner is usually larger than the increase in waiting time.
 For a weakly persistent bath, it is better to keep the waiting time smaller.
 This requires the IHP to also take a leap of faith if the travel direction is not ideal (larger GO angle), given the possibility to opt out again with the next decision.
 Second, the circular motion inherent of cABPs is reflected by the anticipation angle, opposing the circular frequency.

Our minimal model assumes a non-reciprocal coupling between the IHP and the bath particles.
While effective non-reciprocal interactions are ubiquitous in nature and are known to generate intriguing non-equilibrium behaviour \cite{ivlev2015statistical,agudo2019active,you2020nonreciprocity,levis2020flocking,mugica2022scale,gomez2022intermittent},
it would also be interesting to model and investigate the reciprocal effects exerted by the IHP on the bath for various reasons.
First, such a  coupling can also lead to unexpected joint motion of the composite particle.
For example, it has been shown that an ABP reacts to its passive cargo by changing its typical dynamics in an activity gradient \cite{vuijk2021chemotaxis,muzzeddu2023taxis}.
Second, alternative types of models could describe the IHP as a carrier vesicle \cite{kokot2022spontaneous,uplap2023design, wittmann2023statistics}, which learns to
selectively uptake and release active particles to get a directional net push in its interior.
 Third, an intriguing philosophical perspective would be to assess the thermodynamic role played by the IHP.
One might argue that an IHP  actually extracting its required work from the active bath
 effectively learns how to act as a Maxwell's demon.

Our model can be generalised in various other ways for future studies.
One important step to go beyond the idealised systems considered here is to include more realistic  bath particles interacting with each other.
The need for anticipating the emerging collective behaviour, which can lead to complex nonequilibrium patterns such as clusters \cite{theurkauff2012dynamic} and vortexes \cite{wensink2012meso},
would challenge simple learning strategies.
One interesting application is the possible transfer of the acquired knowledge to macroscopic bodies such as robots \cite{zhao2022twisting,kamp2024robots}, which shall perform comparable navigational tasks.
As such, an important aspect that should be taken into account is the role of inertia \cite{lowen2020inertial}, which brings about a memory effect of the past trajectories
and might also require more sophisticated learning tools.

\acknowledgments

We would like to thank Alexander R.\ Sprenger, Lorenzo Caprini and Rahul Gupta for stimulating discussions.
Funding by the Deutsche Forschungsgemeinschaft (DFG) through the SPP 2265 under grant numbers LO 418/25-2 (HL) and WI 5527/1-2 (RW) is gratefully acknowledged.

\appendix

\section{Theory for the MSD of the $Q$-learning IHP \label{app_MSDTheory}}

While the MSD of the bath particles is exactly known \cite{teeffelen2008loewen},
 there are also specific rules for hitchhiking.
We can thus provide close analytical estimates for the MSD of the IHP,
where we focus on the $Q$-learning IHP in a bath of straightly swimming ABPs.
Our considerations rest on two pillars.
First, if the IHP has a suitable travel partner, its motion alongside the persistent trajectory of this ABP also occurs with self-propulsion velocity $v_0$ and is ballistic.
Second, as the IHP may also be at rest, the effective hitchhiking velocity $v_\text{H}(t)$ becomes an average quantity
which depends on time-dependent hitchhiking probability $p_\text{H}(t)$, i.e., the probability that \textit{the IHP is currently attached to an ABP}.
Such an ABP is chosen for hitchhiking if
(i) it is within scan range $R_\text{s}$ (at a time given by a multiple of $\tau_Q$) when a decision is made,
(ii) it is closest to the IHP such that is actually selected as a potential vessel (in case more than one ABPs are within scan range) and
(iii) its orientation $\phi_i\in\Phi_\text{GO}$  is considered favourable.
Hence, the bath density $\rho$ is a crucial control parameter.
Moreover, after the IHP has found a suitable ABP, conditions (i) and (ii) will be automatically fulfilled by the same ABP when the next decision is made, while condition (iii) needs to be evaluated again.
Because of the possible choice of the IHP to leave a travel partner that runs out of persistence, we eventually also need to take into account the persistence time $\tau$ of the bath particles, which will indirectly influence the hitchhiking probability $p_\text{H}(t)$.

If we know the hitchhiking probability $p_\text{H}(t)$, we can determine the effective hitchhiking velocity
\begin{equation}
    v_\text{H}(t) = p_\text{H}(t) v_0\,
    \label{eq:vt}
\end{equation}
right away, such that the resulting displacement follows as
\begin{equation}
    x(t) = \int_0^t v_\text{H}(t') \mathrm{d}t'\,.
    \label{eq:xt}
\end{equation}
As a directed IHP motion is ensured by its training and the resulting selective choice of travel partners, we conveniently assume here an effective one-dimensional displacement,
which directly yields the desired MSD$=x^2(t)$ by squaring the result.
In what follows, we discuss the form of $p_\text{H}(t)$ in three steps.

As a first step, we want to calculate the probability $p(t)$ that \textit{the IHP has found its first travel partner before or at time $t$} as an upper bound to $p_\text{H}(t)$.
The first decision is made at $t=0$, where the initial probability $p(0)$ can be stated as \textit{there is at least one ABP within range and the selected ABP has a favourable orientation}.
In an infinite noninteracting bath of density $\rho$, the probability that $k$ particles are found within the area $R_\text{s}^2\pi$ (surrounding the IHP) is given by the Poisson distribution
\begin{equation}
    P_k(\lambda)=\frac{\lambda^k}{k!}e^{-\lambda}\,,
    \label{eq:poisson}
\end{equation}
where $\lambda=\rho R_\text{s}^2\pi$ represents the average number of particles.
Together with the GO-angle $\Delta\phi$ in the definition of $\Phi_\text{GO}$ from Eq.~\eqref{eq_PhiGO},
we find the required initial probability as
\begin{equation}
    p(0) = (1-P_0(\rho R_\text{s}^2\pi)) \,\frac{\Delta\phi}{2\pi}= \left(1-e^{-\rho R_\text{s}^2\pi}\right)\frac{\Delta\phi}{2\pi}\,.
    \label{eq:p0Poisson}
\end{equation}
For the sake of comparing predictions based on $p(0)$ to our simulation results,
we should take into account the finite periodic simulation box of length $L$,
such that the binomial distribution should be used instead of the Poisson distribution.
Accordingly, the result analogous to Eq.~\eqref{eq:p0Poisson} is given by Eq.~\eqref{eq:p0}, as stated and discussed in the main text and shown in Fig.~\ref{fig_MSD}b.
 For sufficiently large $L$, both expressions become equivalent.
Irrespective of how $p(0)$ is specified, we can obtain the probability
\begin{equation}
    p(n) = \sum_{m=0}^n\left(1-p(0)\right)^m p(0)
    \label{eq:pH}
\end{equation}
that \textit{the journey of the IHP has started no later than with the $n+1$st decision} by cumulating the probabilities $\left(1-p(0)\right)^m p(0)$ that \textit{exactly the $m+1$st decision results in choosing a travel partner for the first time}.
Recalling that these decisions are made at  $t = n \tau_\text{Q}$ with integers $n\geq0$,
we obtain the desired probability $p(t)$ from Eq.~\eqref{eq:pH} by setting $n=\lfloor t/\tau_Q\rfloor$,
where the brackets evaluate to the nearest integer smaller than or equal to the argument.

 As a second step, we notice that $p(t)$ is not exactly the hitchhiking probability $p_\text{H}(t)$ required in Eq.~\eqref{eq:vt}, because it does not account for the need of the IHP to eventually leave its initial (or current) travel partner and wait again until the next suitable ABP is chosen to continue the trip in the desired direction.
 Instead, we see from the geometric series in Eq.~\eqref{eq:pH} that $p(t)\rightarrow1$ for large times, which merely describes that the IHP will always find a suitable travel partner if it just waits long enough.
While it is indeed possible to derive an explicit formula for  $p_\text{H}(t)$, this result will be impracticable for  evaluation and we only briefly state the basic idea.
The conditional staying probability $p_S(t-t_0)$ that the IHP, that has attached to a bath particle at time $t_0$, will still make a GO decision at time $t$ can be determined from the Gaussian solution of the angular diffusion process in Eq.~\eqref{eq:eq_of_motionPHI} with $\omega=0$, i.e., when the diffusing angle exceeds the limits described by $\Phi_\text{GO}$.
The most notable insight of this exercise is that the relevant change in this probability occurs on a time scale given by $\tau$, which we will refer to below.
The appropriate calculation of $p_\text{H}(t)$ would then require for an iterative correction of
Eq.~\eqref{eq:pH} by introducing factors of the form $p_S(m\tau_Q)$,
accounting for the staying probability after $m$ decisions, which remains nonzero for all decisions after first choosing a travel partner.
Likewise, with probability $1-p_\text{s}$, the following decision must already account for the possibility of attaching to a new travel partner and so on.
This means, that a new term of the form of Eq.~\eqref{eq:pH} must be added after each two decisions, eventually leading to an infinite number of sums.

As a third step, we rather seek here for compact expressions to approximate $p_\text{H}(t)$ only using the available result for $p(t)$, which certainly captures the correct behaviour in the early stages of the IHP dynamics.
For later times, we utilise the insight that the wait--GO--NO GO sequence described in the previous paragraph will roughly repeat itself after a time span given by $\tau$.
Thus we average our result for $p(t)$ over one persistence time to
 estimate long-time limit $\bar{p}_\infty:=p_\text{H} $ the hitchhiking probability as
\begin{equation}
    \bar{p}_\infty \approx \frac{1}{\tau} \int_0^\tau p(t) \mathrm{d}t\,.
    \label{eq:v_bar}
\end{equation}
This integral can be expressed as an analytic series by summation over $p(n)$ Eq.~\eqref{eq:pH}.
In summary, we approximate the overall hitchhiking probability as
\begin{equation}
    p_\text{H}(t) \approx \begin{cases}
     p(t)\,, &  t \lessapprox \tau \\
     \bar{p}_\infty\,, &  t \gg \tau \\
   \end{cases}\,
    \label{eq:hitchhiking speed}
\end{equation}
and, accordingly, the effective hitchhiking velocity $v_\text{H}(t)$ using Eq.~\eqref{eq:vt}.
These two cases provide an excellent description of the IHP's MSD in the respective limits.
 Specifically, we show in Fig.~\ref{fig_MSD} the short-time approximation
\begin{equation}
    \text{MSD}_0 = \left(v_0\int_0^t p(t') \mathrm{d}t'\right)^2\,.
    \label{eq_MSD0}
\end{equation}
and the long-time asymptote
\begin{equation}
    \text{MSD}_\infty = v_0^2\bar{p}_\infty^2t^2\,.
    \label{eq_MSDi}
\end{equation}

\section{Initial hitchhiking probability of the potential IHP \label{app_MSDTheoryPOT}}

To get a feeling for the performance of the potential IHP relative to the $Q$-learning IHP, we compare the initial hitchhiking probabilities of both models.
For the potential IHP this probability, which we denote by $\tilde{p}(0)$, can be stated as \textit{there is at least one ABP within range and at least one of them has a favourable orientation}.
For the reason of obtaining a compact representation, we work with the Poisson distribution from Eq.~\eqref{eq:poisson}, appropriate for a sufficiently large system.
Doing so, we find
\begin{equation}
\tilde{p}(0)=1-\sum_{k=0}^\infty P_k(\rho R_\text{s}^2\pi) \left(1-\frac{\Delta\phi}{2\pi}\right)^k
=1-e^{-\rho R_\text{s}^2\pi\frac{\Delta\phi}{2\pi}}
\end{equation}
as the complement of having each $k$ bath particles within scan radius, while all of them have an unfavourable orientation.
Comparing this result to Eq.~\eqref{eq:p0Poisson}, we see that both expressions have the same low-density limit
\begin{equation}
p(0)=\rho R_\text{s}^2\pi\frac{\Delta\phi}{2\pi}+\mathcal{O}(\rho^2)=\tilde{p}(0)\,,
\end{equation}
which explicitly depends on the scan radius $R_\text{s}$.
In general, we see that $\tilde{p}(0)$ is larger than $p(0)$, as can be shown explicitly from the ratio
\begin{equation}
\frac{p(0)}{\tilde{p}(0)}=1-\frac{\rho R_\text{s}^2\pi}{2}\left(1-\frac{\Delta\phi}{2\pi}\right)+\mathcal{O}(\rho^2)\,.
\end{equation}
Specifically, taking the high-density limits
\begin{equation}
\lim_{\rho\rightarrow\infty}p(0)=\frac{\Delta\phi}{2\pi}<1\,,\ \ \ \lim_{\rho\rightarrow\infty}\tilde{p}(0)=1\,,
\end{equation}
the difference between the two models becomes obvious.
As the presence of one bath particle with favourable orientation is sufficient for the potential IHP, there surely is a hitchhiking possibility in the high-density limit.
In contrast, the $Q$-learning IHP only has the chance to select a favourable bath particle with a probability determined by the fraction of available GO-angles,
despite the plethora of suitable travel partners.

%\bibliography{references.bib}


\end{document}